\documentclass[a4paper,epjCONF,columns]{svjour} 
\usepackage{graphicx}  
\usepackage[varg]{txfonts} 
\usepackage[latin1]{inputenc}
\session-title{2011 Hadron Collider Physics symposium (HCP-2011)}
\begin{document}
\title{Quarkonium production at ATLAS}
\author{Darren~D.~Price\thanks{\email{Darren.Price@cern.ch}} (on behalf of the ATLAS collaboration)}
\institute{Department of Physics, Indiana University, Bloomington, IN 47405, USA.}
\abstract{
The production of quarkonium is an important testing ground for QCD calculations. The $J/\psi$ and $\Upsilon$ production cross-sections
are measured in proton-proton collisions at a centre-of-mass energy of 7~TeV with the ATLAS detector at the LHC. Differential cross-sections are presented as a 
function of transverse momentum and rapidity. The fraction of $J/\psi$ produced in B-hadron decays is also measured and the 
differential cross-sections of prompt and non-prompt $J/\psi$ production determined separately. Measurements of the fiducial production cross-section of the $\Upsilon\textrm{(1S)}$ 
and observation of the $\chi_{c,bJ}$ states are also discussed.}
\maketitle
\section{Introduction}
\label{intro}

Quarkonia are formed from a quark pair of same flavour and should represent one of the simplest systems described by QCD theory. 
Heavy quarkonium is a multiscale system, allowing for rigorous tests of the interplay between perturbative and non- perturbative QCD and a 
rich spectrum of radial and orbital excitations allow studies of spectroscopy and decay dynamics\,\cite{QWGReviews}. In addition, quarkonia is an ideal probe of cold and 
hot nuclear matter effects\,\cite{JpsiSuppression}. Many open questions exist in this area, and it has proven difficult thus far to universally describe the kinematics 
and production properties of quarkonia. ATLAS plans to investigate quarkonium production through measurements of production, spin-alignment and 
associated hadronic activity of these various states. The first production measurements toward this goal are outlined below.

\section{Inclusive, prompt and non-prompt {\boldmath $J/\psi$} production cross-section measurements}

Cross-section measurements presented here make use of data collected via a single muon trigger, first with no threshold on $p_T$ then later with a threshold at 4~GeV
as instantanous luminosity increased. Trigger and reconstruction efficiencies are measured in data and validated with Monte Carlo simulations.
Weights incorporating acceptance and efficiency corrections are applied to quarkonia candidates on an event-by-event basis before fits are used to extract a cross-section.

The acceptance corrections represent the probability for quarkonium with given kinematics to pass basic selection cuts. The exact value of this probability in a given phase
space is dependent on nature of the as yet unmeasured spin alignment of quarkonium.
Five spin-alignment scenarios have been identified that induce the largest envelope of variation on visible cross-sections. 
Measurements are repeated under application of different acceptance maps reflecting different polarisation states as a systematic effect on measured production observables.

Production of $J/\psi$ can occur promptly from the hard interaction, or may be produced non-promptly via decay of a B-hadron.
$J/\psi$ from b-decays have positive displaced di-muon vertices and can be distinguished from prompt production (and non-$J/\psi$ backgrounds) via the pseudo-proper time discriminant 
$\tau=L_{xy}\cdot m^{J/\psi}_{\textrm{PDG}}/p^{J/\psi}_{T}$, where $L_{xy}$ is the transverse decay length of the $J/\psi$ vertex.

\begin{figure}[htbp]
  \includegraphics[width=0.5\textwidth]{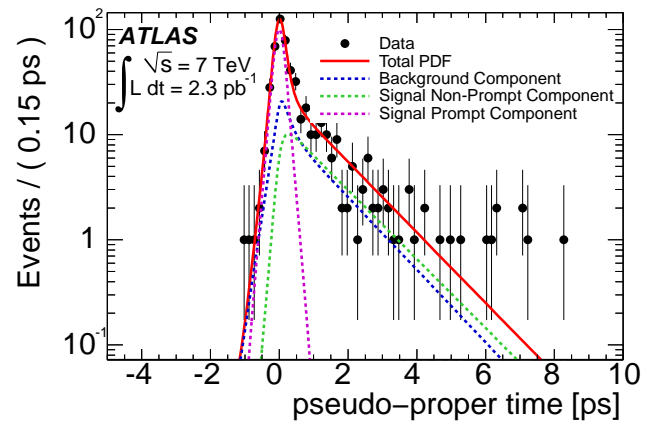}
  \caption{Pseudo-proper time distribution of $J/\psi\to\mu^+\mu^-$ candidates in the signal mass region, for $9.5 < p_T(J/\psi) < 10.0$~GeV and $|y^{J/\psi}|<0.75$. 
    The points represent measured data, the solid line is the result of an unbinned maximum likelihood fit to all di-muon pairs in the $2.5-3.5$~GeV mass region 
    projected onto the pseudo-proper lifetime distribution.}
  \label{fig:lifetime}
\end{figure}

\begin{figure}[htbp]
  \includegraphics[width=0.5\textwidth]{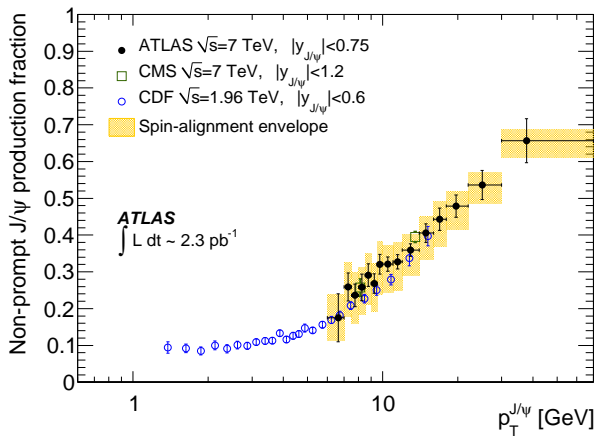}
  \caption{Production fraction of non-prompt to inclusive $J/\psi$ as a function of $J/\psi$ transverse momentum.
    Overlaid is a band representing the maximal variation of the result under various spin-alignment scenarios.
    Comparison is made to existing CDF and CMS results.}
  \label{fig:fraction}
\end{figure}

A simultaneous unbinned maximum likelihood fit to invariant mass and lifetime allows us to distinguish prompt and non-prompt $J/\psi$ production
from combinatorial background (see Figure~\ref{fig:lifetime} for an example of such a fit projected onto the lifetime distribution) 
and determine the fraction of $J/\psi$ produced via $B$-decays as a function of $p_{T}$ and rapidity (results for one rapidity slice are shown in Figure~\ref{fig:fraction}). 

The data analysed allow for study of the $B$-production fraction across a $p_T$ range of 6 to 70~GeV, to far higher transverse momenta than have previously been studied.
A strong dependence of the fraction is observed as a function of $p_{T}$, but slow dependence on rapidity in the range ($0<|y|<2.4$) of rapidities studied. 
Good agreement with CDF data ($p\overline{p}$ collisions at $\sqrt{s}=1.96$~TeV) is observed within the experimental uncertainties, suggesting the $B$-fraction has limited dependence
on centre-of-mass energy and initial colliding particles, particularly at large $p_T$.

\begin{figure}[htbp]
  \includegraphics[width=0.5\textwidth]{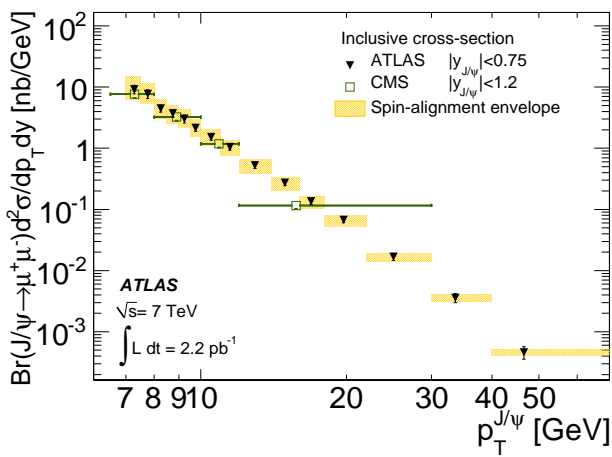}
  \caption{Inclusive $J/\psi$ production cross-section as a function of $J/\psi$ transverse momentum in the $|y|<0.75$ rapidity bin.} 
  \label{fig:inclusive}
\end{figure}

Applying candidate-by-candidate efficiency and acceptance weights to di-muon pairs in $p_T-y$ bins, extract signal yield from unbinned maximum likelihood fit to $J/\psi$ peak. 
Figure~\ref{fig:inclusive} shows an example of an inclusive differential cross-section extracted for one rapidity bin (four in total), as a function of $J/\psi$ $p_T$.

By combining the information from the inclusive cross-section and $B$-fraction measurements, one can extract a non-prompt ($J/\psi$ from $b$-decays) and 
prompt $J/\psi$ cross-section versus $p_T$ and rapidity.
\begin{figure}[htbp]
  \includegraphics[width=0.5\textwidth]{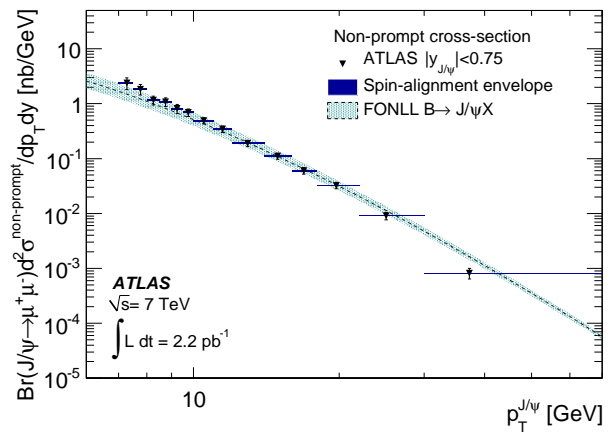}
  \includegraphics[width=0.5\textwidth]{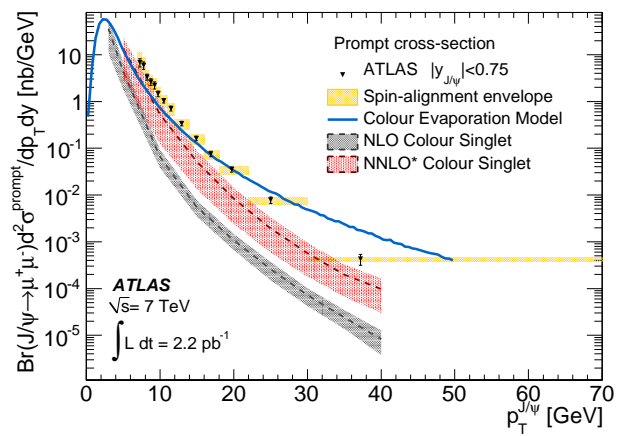}
  \caption{Non-prompt (top) and prompt (bottom) $J/\psi$ production cross-sections as a function of $J/\psi$ transverse momentum, compared to theoretical predictions (FONLL, in the case of
    non-prompt production; NLO and NNLO* and Colour Evaporation Model predictions for prompt production). 
    Overlaid is a band representing the variation of the result under various spin-alignment assumptions on the non-prompt and prompt components. 
    The central value assumes an isotropic polarisation for both prompt and non-prompt production. }
  \label{fig:promptnonpromp}
\end{figure}
Measurements of $J/\psi$ from $b$-decays is found to agree well with Fixed-Order Next-to-Leading-Log theoretical predictions. Comparisons to colour singlet NNLO* 
pQCD predictions and the phenomenological Colour Evaporation Model show significant discrepancies in shape and normalisation, highlighting the uncertainties
that exist in explaining the nature of prompt $J/\psi$ production.

\section{Prompt {\boldmath $\Upsilon(1S)$} fiducial production cross-section measurement}

Measurement of the $\Upsilon(1S)$ cross-section has been conducted in a similar manner as with $J/\psi$, also using single muon triggers. 
In contrast to the $J/\psi$ results, this measurement is presented in a fiducial region: $p^{\mu}_{T}>4$~GeV, $|\eta^{\mu}|<2.5$ to remove 
the uncertainty due to spin alignment in the cross-section measurement.
%

Unfolded differential cross-sections are compared to NLO pQCD and significant disagreement is noted.
This behaviour is consistent with other theoretical predictions\,\cite{Lansberg} at NLO when compared to prompt $J/\psi$ and $\Upsilon$ data at the Tevatron and LHC,
highlighting the need for additional higher order contributions to account for the observed production.
\begin{figure}[htbp]
  \includegraphics[width=0.48\textwidth]{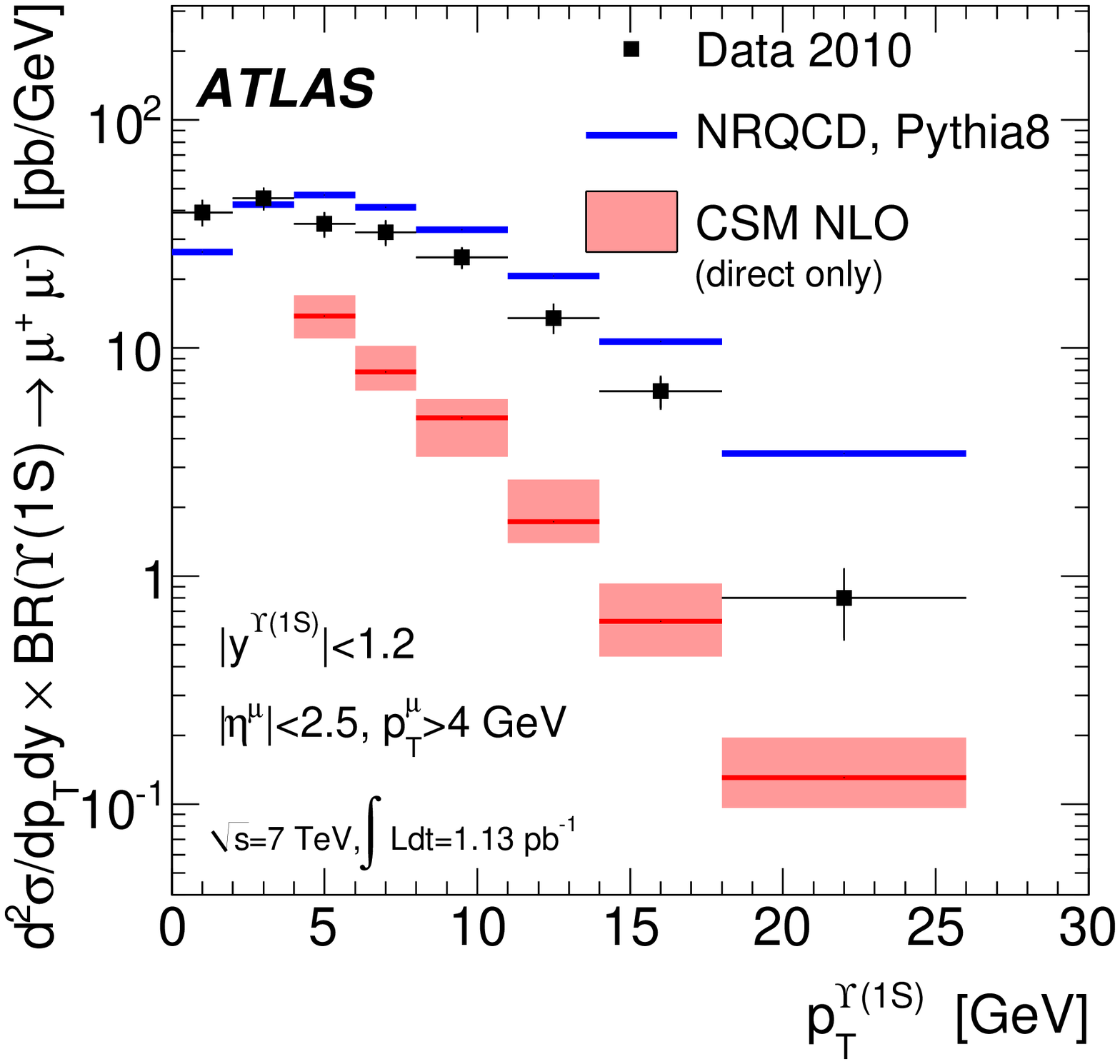}
  \includegraphics[width=0.48\textwidth]{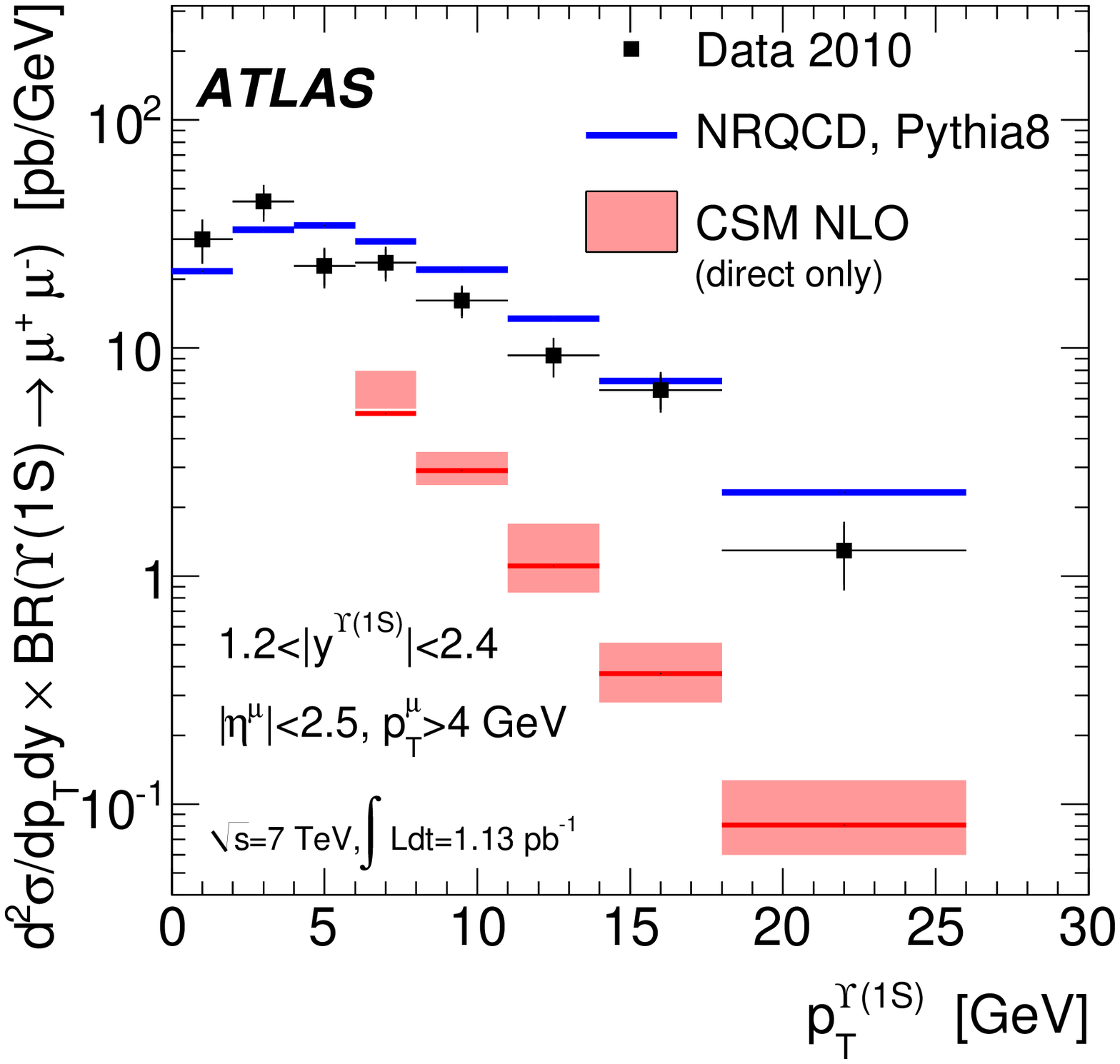}
  \caption{Differential prompt $\Upsilon(1S)\to\mu^+\mu^-$ production cross-section for $|y^{\Upsilon(1S)}|<1.2$ (top) and $1.2<|y^{\Upsilon(1S)}|<2.4$ (bottom) 
    as function of $p_{T}^{\Upsilon(1S)}$ for $p^{\mu}_{T}>4$~GeV, $|\eta^{\mu}|<2.5$. Comparison is made to colour-singlet NLO {\it direct} production predictions, and
    the shaded area shows the change in the theoretical prediction when varying the renormalisation and factorisation scales by a factor of two. 
    A comparison is also made to a model based on NRQCD native to \textsc{Pythia8}, for a particular choice of parameters.}
  \label{fig:Upsilon}
\end{figure}

\section{Reconstruction of {\boldmath $\chi_{cJ}$} mesons through radiative decays}

The $\chi_{c1}$ and $\chi_{c2}$ mesons in radiative decays to a $J/\psi$ and a photon have been observed via calorimetry. 
Measuring the production cross-sections of the $\chi_{cJ}$ are crucial for precise understanding of $J/\psi$ production and acts as a test of pQCD in its own right.

\begin{figure}[htbp]
  \includegraphics[width=0.5\textwidth]{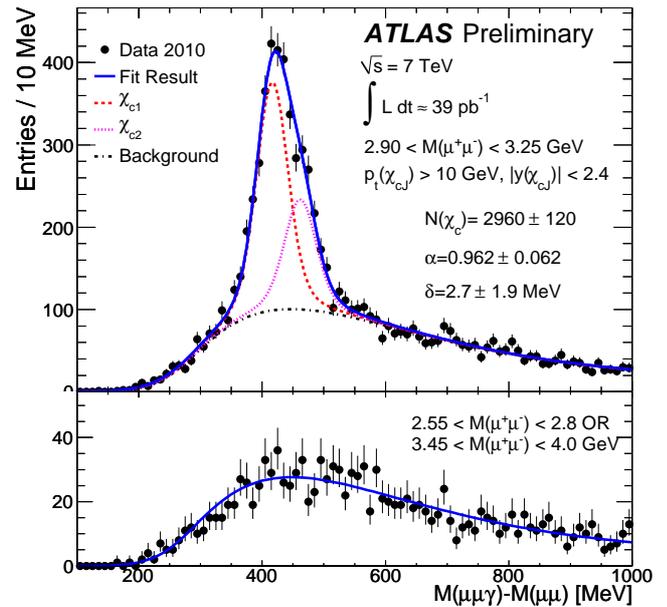}
  \caption{The results of a simultaneous fit to the signal region (top) and background ($J/\psi$ sideband) region (bottom), showing 
    signals of $\chi_{c1}$ and $\chi_{c2}$ mesons reconstructed (via calorimetry measurement) through their radiative decays to $J/\psi+\gamma$,
    viewed as a function of the invariant mass difference between the $\mu\mu\gamma$ and $\mu\mu$ systems. 
  }

  \label{fig:chic}
\end{figure}

Photons at ATLAS can also be identified and measured through reconstruction of the tracks from electron-position pairs associated with a photon conversion occurring in the Inner Detector.
The tracking capabilities of ATLAS allow for much more precise measurement of the photon kinematics through this method, but is balanced by a low probability for a conversion to occur and
successfully be reconstructed. Measurement via calorimetric techniques (as shown in Figure~\ref{fig:chic}) benefit from increased signal efficiency
at a cost of reduced resolution. Recently, similar techniques have been used by ATLAS to reconstruct the $\chi_{bJ}(mP)$ states (via $\Upsilon(nS)+\gamma$ decays)
in both calorimetry and tracking modes and has led to the discovery\,\cite{chib} of a new set of states, the $\chi_{bJ}(3P)$.

\begin{thebibliography}{}

\bibitem{QWGReviews}
  N.~Brambilla {\it et al.}  [Quarkonium Working Group Collaboration],
  hep-ph/0412158;
  N.~Brambilla {\it et al.},
  Eur.\ Phys.\ J.\ C {\bf 71} (2011) 1534
  [arXiv:1010.5827 [hep-ph]].

\bibitem{JpsiSuppression}
  The ATLAS Collaboration,
  Phys.\ Lett.\ B {\bf 697} (2011) 294
  [arXiv:1012.5419 [hep-ex]].

\bibitem{Jpsi}
  The ATLAS Collaboration,
  Nucl.\ Phys.\ B {\bf 850} (2011) 387
  [arXiv:1104.3038 [hep-ex]].

\bibitem{CDFjpsi}
  D.~Acosta {\it et al.}  [CDF Collaboration],
  Phys.\ Rev.\ D {\bf 71} (2005) 032001
  [hep-ex/0412071].

\bibitem{CMSjpsi}
  V.~Khachatryan {\it et al.}  [CMS Collaboration],
  Eur.\ Phys.\ J.\ C {\bf 71} (2011) 1575
  [arXiv:1011.4193 [hep-ex]].

\bibitem{Upsilon} 
  The ATLAS Collaboration,
  Phys.\ Lett.\ B {\bf 705}, 9 (2011)
  [arXiv:1106.5325 [hep-ex]].

\bibitem{Lansberg}
  J.~P.~Lansberg,
  Eur.\ Phys.\ J.\ C {\bf 61} (2009) 693
  [arXiv:0811.4005 [hep-ph]].

\bibitem{chic}
  The ATLAS Collaboration,
  ATLAS-CONF-2011-136 (https://cdsweb.cern.ch/record/1383839).

\bibitem{chib}
  The ATLAS Collaboration,
  arXiv:1112.5154 [hep-ex].

\end{thebibliography}
\end{document}